# Hybrid Pixel Detector Development for the Linear Collider Vertex Detector

M.Battaglia[1], S.Borghi[2], M.Caccia[3], R.Campagnolo[2], W.Kucewicz[4], H.Palka[5], A.Zalewska[5]

[1] University of Helsinki, P.O.Box 9, 00014 Helsinki, Finland
[2] Università degli Studi di Milano, Dipartimento di Fisica, via Celoria 16, 20133 Milano, Italy
[3] Università degli Studi dell'Insubria, Dipartimento di Scienze, Via Valleggio 11, Como, Italy
[4] University of Mining and Metallurgy, al. Mickiewicza 30, 30-059 Kraków
[5] Institute of Nuclear Physics, ul. Kawiory 26 A, 30-055 Kraków

*Abstract*

In order to fully exploit the physics potential of future high energy $e^+e^-$ linear collider, a Vertex Detector providing high resolution track reconstruction is required. Hybrid silicon pixel detectors are an attractive option for the sensor technology due to their read-out speed and radiation hardness but have been so far limited by the achievable single point resolution. A novel layout of hybrid pixel sensor with interleaved cells to improve the spatial resolution has been developed. The characterisation of the first processed prototypes is reported.

## I. INTRODUCTION

High energy experiments at an electron-positron linear collider, operating at centre-of-mass energies up to 1 TeV, are one of the milestones in the future of Particle Physics. The linear collider will complement the physics reach of the Tevatron at FERMILAB and the Large Hadron Collider at CERN in the study of the mechanism of electro-weak symmetry breaking and in the search for new physics beyond the Standard Model. Both precision measurements and particle searches set stringent requirements on the efficiency and purity of the flavour identification of hadronic jets since final states including short-lived *b* and *c*-quarks and $\tau$ leptons are expected to be the main signatures. The main characteristics of the Vertex Detector may be summarised as follows:

- Single point resolution and detector thickness. The jet flavour tagging requirements can be turned into a track impact parameter resolution, required to be $5\mu m \oplus \frac{10\mu m}{p \cdot sin^{3/2}(\theta)}$ in the plane orthogonal to the colliding beams, where p is the particle momentum and $\theta$ is the polar angle with respect to the beam axis; the first term is directly connected to the detector single point resolution and geometry and the second term accounts for the multiple scattering. Assuming the first sensitive layer at 12 mm radius from the interaction point and the outermost at 100 mm radius, the desired impact parameter performances may be achieved with a single point resolution of 7 µm and a material budget of 0.5% $X_o$ for each layer.

- Time stamping. Single bunch crossing will occurr every 330 ns in case of the TESLA Linear Collider project and its identification will reduce pair and $\gamma\gamma$ background to 0.1 hits/mm$^2$.

- Detector granularity. A sensitive cell area below 150 x 150 µm$^2$ is needed to keep the occupancy from pairs and hadronic jets below 1%.

At the moment, two silicon sensor technologies already used at collider experiments have the potential to satisfy these specifications: the Charged Coupled Devices (CCD) and hybrid pixels sensors. Hybrid pixel sensors have the advantage of allowing fast time stamping and sparse data scan read-out, thereby reducing the occupancies due to backgrounds, and of being tolerant to neutron fluxes well beyond those expected at the linear collider. Both these characteristics have been demonstrated for their application in the LHC experiments. On the other hand, the improvement of the pixel sensor spatial resolution and the reduction of its total thickness represent areas of needed R&D that are specific to the linear collider application. CCD require a complementary development: their resolution and thickness is already quite close to what is needed; on the other hand, the readout speed and the radiation tolerance has to be improved.

Single point precision at the 5 µm level can be obtained by sampling the diffusion of the carriers generated along the particle path and adopting an analog read-out to interpolate the signals of neighbouring cells. Since the charge diffusion r.m.s. is ~ 8 µm in 300 µm thick silicon, an efficient sampling requires a pixel pitch well below 50 µm. At present, the most advanced pixel read-out electronics have minimum cell dimensions of 50 x 300 µm$^2$, limiting an efficient signal interpolation. Even if a sizeable reduction in the cell dimension may be envisaged by the deep submicron trend in the VLSI development, a sensor design overcoming such basic limitation is definitely worth being explored

## II. FIRST PROTOTYPE PRODUCTION

The pixel detector design discussed here exploits a layout inherited from the microstrip detectors [1] where it is assumed to have a read-out pitch *n* times larger than the pixel pitch. The proposed sensor layout is shown in fig.1 for *n*=4. In such a configuration, the charge carriers created underneath an interleaved pixel will induce a signal on the output nodes, capacitively coupled to the interleaved pixel. In a simplified model where the detector is reduced to a capacitive network, the ratio of the signal amplitudes on the output nodes at the left hand side and right hand side of the interleaved pixel (in both dimensions) should have a linear dependence on the particle position. The ratio between the inter-pixel capacitance and the pixel capacitance to backplane plays a crucial rule in the detector design, as it defines the signal amplitude reduction (an effective charge loss) at the output nodes and at last the sustainable number of interleaved pixels. Recent results on 200 µm read-out pitch microstrip detectors have been published, and a 10 µm resolution has been achieved in a layout with 3 interleaved strips (50 µm strip pitch) and for a S/N ≈ 80 [2]. Similar results may be expected in a pixel detector, taking into account a lower noise is achievable because of the intrinsically smaller load capacitance and the charge is possibly shared on four output nodes reconstructing the particle position in two dimensions. Improvements are certainly possible sampling the diffusion with a smaller pitch.

Prototypes of detectors with interleaved pixels have been designed and manufactured. The layout of one of the structures is shown in fig. 1. A series of guard rings defines the detector sensitive area. A bias grid allows the polarisation of the interleaved pixels too; each $p^+$ implant is connected to the metal bias line by polysilicon resistor in the 1-3 MΩ range. A metal layer is deposited on top of pixels to be connected to the VLSI cell. The back plane has a meshed metal layer to allow the use of an infrared radiation diode for charge collection studies. In a 4" wafer 36 structures were fit, for 17 different layouts ; a VLSI cell of 200 x 200µm$^2$ or 300 x 300 µm$^2$ was assumed and detectors with a number of interleaved pixels ranging between 0 and 3 and different areas were designed.

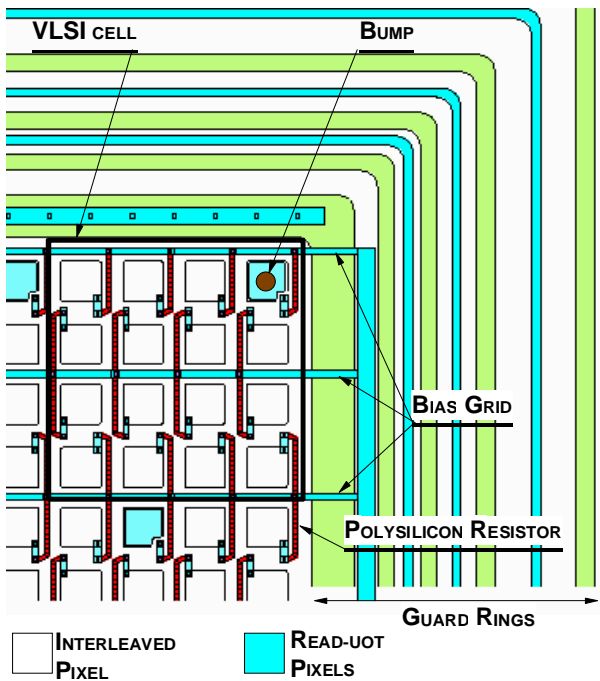

Figure 1: Layout of the corner of the pixel detector test structure with 50 µm implant and 200 µm read-out pitch; the other structures differ by the pitch.

The high resistivity wafers (5 –8 kΩcm) were processed together with an equal number of low resistivity wafers for process control, the details of which have been outlined in [3]. Two wafers were retained by the factory for a destructive analysis and two others were stored for later use. All of the structures on five wafers were visually inspected, tested up to 250 V and characteristics I-V and C-V curves produced. The typical behaviour of a good structure is shown in fig.2 [4].

a)
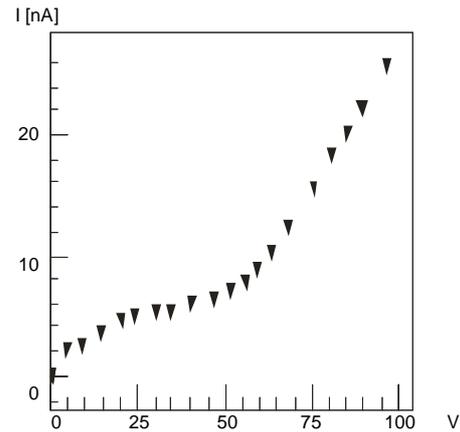

b)
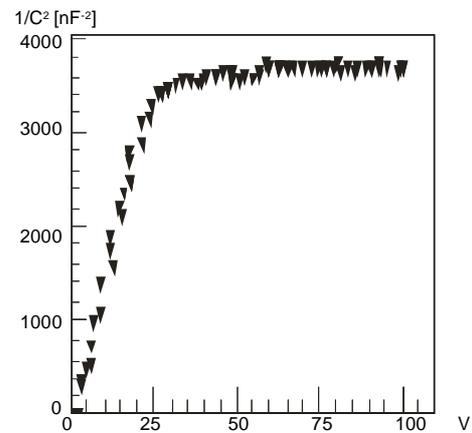

Figure 2: I vs. V (a) and $1/C^2$ vs. V (b) characteristics.

. In fig. 3 the values of the currents and detector capacitancies at depletion voltage are shown for all of the structures in one of the wafers. Neither the design nor the technology show any fundamental flaw: the leakage current for good structures is at the 10 nA/cm$^2$ level; breakdown occurs well beyond the full depletion voltage, achieved at the expected value according to the wafer thickness and the Silicon resistivity; the full area detector capacitance matches the calculated values. The average yield over the tested wafers is about 50%; rejected structures feature an extremely high leakage current independent of the applied voltage, correlated to surface defects possibly associated to plasma etching of the Al pattern. An improvement in the technology sequence, requiring a better planarisation and avoiding plasma etching has been agreed with the manufacturer.

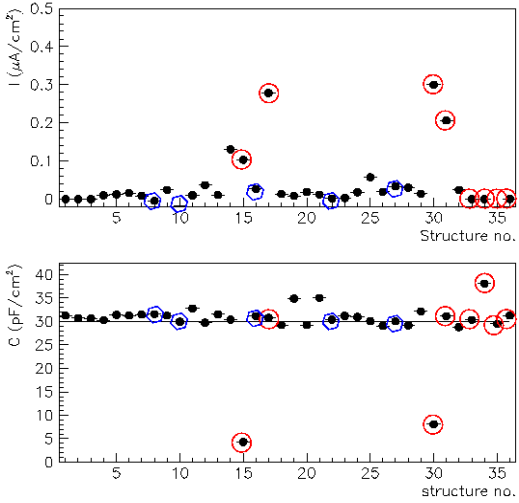

Figure 3: Leakage current and capacitance at full depletion voltage for all 36 structures in one of the wafers; circles define the detector rejected because of breakdown voltage below 100 V; hexagons identify rejected detectors because of interrupted metal lines. The line corresponds to the expected capacitance per unit area.

## III. INTERPIXEL AND BACKPLANE CAPACITANCE MEASUREMENTS

Inter-pixel and backplane capacitances are crucial for the proposed detector design as they determine the effective charge loss and at the sustainable number of interleaved pixels. Since single pixel capacitances are expected to be at the 10 fF level, the measurement conditions were optimised as follows:

- The capacitance was measured using a Hewlett-Packard 4280 CV meter, operating at 1MHz. A cable correction procedure accounting for finite admittances to ground was used. Moreover, an open circuit correction was applied, measuring the capacitance with raised probe tips.
- All of the pixels along a single metal bias line were set in parallel, isolating the line from the bias grid (see fig.1). The number of pixels along a line ranges between 64 and 254, depending on the pixel pitch, thus increasing the measured capacitance values to 0.5-2 pF
- For every structure, 3 bias lines were isolated and the capacitance for each of them, for 3 doublets in parallel and for 1 triplet was measured. The offset in the linear regression for the capacitance versus the number of lines measured the left-over parasitic contributions after the cable correction.
- For each structure, independent measurements of the total interpixel capacitance and of the sum of the interpixel and backplane capacitance were done.

The results are summarised in table 1, where a comparison with a numerical estimate is also shown. The capacitances were calculated solving the Laplace equation inside a 5x5 pixel matrix with a finite element analysis performed using the OPERA-3D package [5].

|  | Chip1 | Chip2 | Chip3 | Chip4 |
|---|---|---|---|---|
| Implant width [μm] | 100 | 60 | 50 | 34 |
| Pixel pitch [μm] | 150 | 100 | 75 | 50 |
| No. pixels in parallel | 64 | 128 | 126 | 254 |
| Measured $C_{interpixel}$ [fF] | 599±5 | 1038±11 | 958±5 | 2098±30 |
| Calculated $C_{interpixel}$ [fF] | 630±60 | 880±67 | 690±70 | 980±40 |
| $C_{backplane}$ [fF] measurement I | 376±40 | 390±50 | 160±45 | 314±35 |
| $C_{backplane}$ [fF] measurement II | 447±5 | 368±5 | 218±5 | 185±5 |
| Calculated $C_{backplane}$ [fF] | 470±70 | 410±90 | 230±35 | 211±60 |

Table 1: Interpixel and backplane capacitance measurements and calculations. The reported values refer to a double column of pixels to the left and right hand side of a single bias line.

The consistency of the measurements is checked comparing the value obtained subtracting the measured interpixel capacitance from the measurement of $C_{interpixel} + C_{backplane}$ (measurement I in the table) and the backplane capacitance from the asymptotic value in the CV curves, normalized by the number of pixels in the matrix (measurement II in table 1). The measurement cross check is satisfactory and the comparison with the calculated values is fair for all of the structures but chip 4, where difficulties both in the measurement and in the simulation were expected because of the small pitch. In particular, the maximum number of elements was preventing an optimal mesh over a larger pixel matrix, crucial for a proper interpixel capacitance evaluation for pixels with 50 μm pitch in a 350 μm thick detector. The numerical estimate is essential to break down the results in the single pixel capacitances to backplane, to the nearest neighbours and to the diagonal and next to nearest neighbours. These values define the characteristics of the detector as a capacitive network and allow a time dependent analysis of the signals induced at the output nodes and an estimate of the charge loss to backplane. The calculated single pixel main capacitances are summarised in table 2.

|  | Chip1 | Chip2 | Chip3 | Chip4 |
|---|---|---|---|---|
| Total $C_{interpixel}$ [fF] | 22.3±1.1 | 12.7±0.6 | 11.8±0.6 | 8.6±0.4 |
| $C_{interpixel}$ to the nearest neighbours [fF] | 3.7±0.2 | 1.9±0.1 | 1.8±0.1 | 1.3±0.1 |
| $C_{backplane}$ [fF] | 7.3±1.1 | 3.2±0.7 | 1.9±0.7 | 0.8±0.2 |

Table 2. Calculated interpixel and backplane capacitances for the single pixel.

According to these values, the charge loss to backplane should never exceed 50%, allowing an efficient signal interpolation irrespective of the particle impact point in the pixel matrix. An experimental measurement on Chip 2 is reported in section IV.

## IV. CHARGE COLLECTION STUDIES

A 4x7 pixel sub-matrix in the prototype structure with 60 μm implant width, 100 μm pixel pitch and 200 μm readout pitch was characterised shining an infrared diode spot with ~80 μm diameter on the detector backplane. The emitted light peaked at a wavelength of 880 nm, with a penetration depth ~ 10 μm [6]. The IR light focusing system was connected to a micrometric (x,y) stage with 1 μm precision along x and 10 μm precision along y, allowing to scan the pixel matrix. The read out pixels were directly connected by wire bonding to the front-end chip used for the readout of the BELLE experiment strip detector. Because of the high IR light intensity, focused on the detector by a multimode fibre, a S/N ~ 100 was achieved even if the noise characteristics were far from being optimal, both because of the chip itself, optimised for a high load capacitance, and because of the wire bonds. Because of the large spot dimensions compared to the pixel implant, testing the structures with 50 μm pixel pitch was not feasible as the diffusion would have completely masked the pure capacitive charge division. Nevertheless, it was possible to extract a meaningful information on the structure with larger pitch.

For each spot position, a statistics of 1000 events was recorded. An initial subset of 300 events was used to initialise the algorithms for the common mode, pedestal and noise calculation. In the second subset of 700 events, the IR pulse was injected every 10 events, allowing for a continuos pedestal tracking. Because of the very limited data volume, no online suppression was active and the data reduction and cluster search was performed off-line. The results summarised below refer to the pulse height on every pixel, averaged over 70 laser pulses.

The charge loss on the backplane is measured by the total pulse height in the cluster as the spot position is changed across the detector, moving from the area underneath a pixel connected to a pre-amplifier to an interleaved junction. The results of the scan are shown in fig. 4.

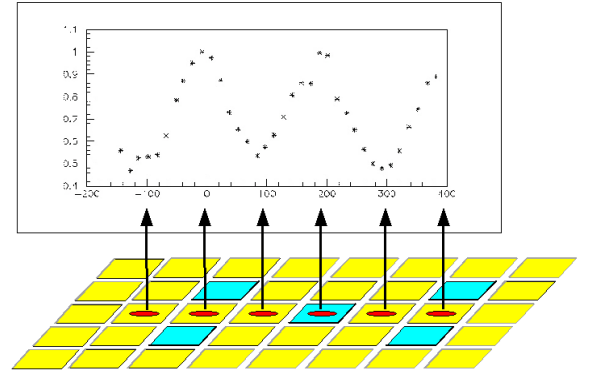

Figure 4. Total pulse height in the cluster, normalised to the maximum value vs. the IR spot position in the pixel matrix. The readout pixels correspond to the dark grey squares.

As expected, the maximum charge loss occurs when the spot is fully contained in the interleaved pixel. The loss corresponds to ~ 50% of the charge, in agreement with what was estimated by a simple network analysis based on the capacitance values reported in section III.

The charge sharing may be described by the η function, defined as $\eta = PH_i/PH_{cluster}$, where $PH_i$ is the pulse height on the reference pixel i, normalised to the cluster pulse height $PH_{cluster}$. The measured distribution is shown in fig.5, where the reference pixel is identified as no. 1 and $PH_{cluster} = PH_1+PH_2+PH_3$ to the left hand side of pixel 1 while $PH_{cluster} = PH_1+PH_4+PH_5$ to its right hand side. The experimental curve can be understood as a superposition of the effects due to the diffusion of the charge carriers created by the IR spot on the neighbouring junctions and by the capacitive charge sharing. The latter may be singled out considering positions of the spot fully contained underneath one pixel and identified by the arrows in fig. 5. The η values for such positions correspond to what can be estimated neglecting diffusion effects and sharing the charge to the output nodes according to a pure capacitive network analysis.

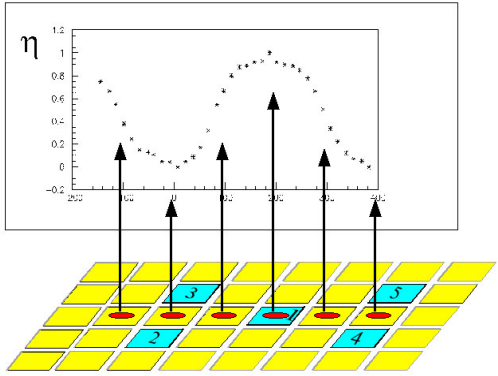

Figure 5. Charge sharing among the different readout pixels, measured by $\eta = PH_1/PH_{cluster}$.

## V. CONCLUSIONS

Prototypes of pixel detectors with interleaved implants between the output nodes were designed and produced. The electrostatics characterisation and preliminary charge collection studies with an Infrared diode confirmed the validity of the detector concept, based on the capacitive coupling among neighbouring pixels. A more refined analysis with a 5 μm light spot and a second prototype production with 25 μm implant pitch, built-in fan out and dedicated structures for interpixel and backplane capacitance measurements are planned on a short time scale.

## VI. AKNOWLEDGEMENTS

The authors wish to thank E.Banas, P.Jalocha and P.Kapusta for their help with the BELLE data acquisition set-up.